\documentclass{ws-procs9x6}

\def\bsbsb{$B_s^0$-$\overline{B}{}_s^0$}
\def\cA{{\cal A}}
\def\Ga{\Gamma}
\def\bb{\overline{B}{}}
\def\dabs{(\Delta a^{B_s}){}}

\newcommand{\refeq}[1]{(\ref{#1})}
\def\etal {{\it et al.}}

\begin{document}

\title{TESTING FOR CPT VIOLATION IN \\
 $B^0_s$ SEMILEPTONIC DECAYS}

\author{R.\ VAN KOOTEN}

\address{Department of Physics, Indiana University,
Bloomington, IN 47405, USA\\
E-mail: rvankoot@indiana.edu}

\author{On behalf of the D\O\ Collaboration%
\footnote{http://www-d0.fnal.gov}}

\begin{abstract}
A D\O\ analysis measuring the charge
asymmetry $A^b_{\mathrm{sl}}$  
of like-sign
dimuon events due to semileptonic $b$-hadron decays
at the Fermilab
Tevatron Collider has shown indications of possible
anomalous CP violation in the mixing of neutral $B$ mesons.
This result has been used to extract the first 
senstivity to CPT violation in the $B^0_s$ system.
An analysis to explore further this anomaly by specifically measuring the semileptonic
charge asymmetry, $a^s_{\mathrm{sl}}$, in $B^0_s$ decays is described, as
well as how a variant of this analysis can be used to
explore a larger set of CPT-violating parameters in the $B^0_s$ system for the
first time.
\end{abstract}

\bodymatter

\section{Introduction}

The interferometric systems of
the particle-antiparticle oscillations of neutral mesons are
particularly sensitive to testing for CP and CPT violation.
In neutral meson systems, the hamiltonian is a $2 \times 2$  matrix relating 
the mass and weak eigenstates.  Mixing between particle and antiparticle
is driven by nonzero off-diagonal matrix elements due to a box 
diagram between $B^0_{(d \thinspace {\mathrm{or}} \thinspace s)}$ and 
$\bar{B}^0_{(d \thinspace {\mathrm{or}} \thinspace s)}$. T (or CP) violation
in mixing can be due to differences between these off-diagonal terms
and results in the two probabilities for oscillation between particle and 
antiparticle not being equal, i.e.,
$P(B^0 \rightarrow \bar{B}^0; t) \neq P(\bar{B}^0 \rightarrow B^0; t)$.
CPT and Lorentz violation involves differences between {\it diagonal} 
terms of this matrix and differences in the probabilities for:
$P(B^0 \rightarrow B^0; t) \neq P(\bar{B}^0 \rightarrow \bar{B}^0; t)$ and can be 
expressed with the parameter\cite{cpt_preprint}
\begin{equation}
\xi = \frac{(M_{11} - M_{22}) - \frac{i}{2}(\Gamma_{11} - \Gamma_{22})}{-\Delta m - \frac{i}{2}\Delta \Gamma} \thinspace \approx \frac{\beta^{\mu} \Delta a_{\mu}}{-\Delta m - \frac{i}{2}\Delta \Gamma},
\label{eq:xidef}
\end{equation}
where $\beta^{\mu} = \gamma(1,\vec{\beta})$ is the 4-velocity of the neutral $B$ meson,
and $\Delta a_{\mu} = r_{q_1} a^{q_1}_{\mu} - r_{q_2} a^{q_2}_{\mu}$ with
$r$ being coefficients with $q_1$ and $q_2$ as meson valence quarks and
$a_{\mu}$ being the constant 4-vector in the Standard-Model Extension langrangian.\cite{SME}
The power of using this system as an inteferometer is that $\Delta m$, i.e., the
mass difference between the heavy and light mass eigenstates, is so tiny,\cite{PDG} i.e., 
for the $B^0_s$ meson, $\Delta m_s = 1.16 \times 10^{-2}$~eV compared with
the typical $B^0_s$ meson mass of approximately 5.4~GeV, allowing sensitivities down to
the  order of $10^{-11}$ on CPT-violating parameters.


\section{D\O\ dimuon charge asymmetry}

The
D\O\ Collaboration has measured\cite{D0_dimuon1,D0_dimuon2} the raw
dimuon charge asymmetry
$A = (N^{\mu^+\mu^+} - N^{\mu^-\mu^-})/(N^{\mu^+\mu^+} + N^{\mu^-\mu^-})$
in $p\bar{p}$ collisions at $\sqrt{s} = 1.96$~TeV
regardless of muon source. From pure physics processes at the
primary interaction, one of the very few sources of same-sign 
dileptons in the same collision event is due to $B$ physics. 
If there is a nonzero asymmetry after correcting for backgrounds,
the assumption is that it is coming from neutral $B$-meson mixing, i.e., 
the dimuon charge asymmetry of semileptonic $B$ decays
$
A^b_{\mathrm{sl}} = (N_b^{\mu^+\mu^+} - N_b^{\mu^-\mu^-})/(N_b^{\mu^+\mu^+} + N_b^{\mu^-\mu^-}).
$
This can occur, for example, when the $b$ quark
decays semileptonically directly
$\overline{B}{}^0_q \rightarrow \mu^-$, but for the $\bar{b}$ quark, there
is first a $B$-meson oscillation before the semileptonic
decay, i.e., $B^0_q \rightarrow \overline{B}{}^0_q \rightarrow \mu^-$. 
Another way to measure this asymmetry is via inclusive `wrong-sign' decays, 
i.e., $\overline{B} \rightarrow \mu^+ X$ which is only possible  through
flavor oscillation of $B^0_d$ and $B^0_s$. A semileptonic charge
asymmetry can then be constructed:
\begin{equation}
a^b_{\mathrm{sl}} =
\frac{\Gamma(\overline{B} \rightarrow \mu^+) - \Gamma(B \rightarrow \mu^- X)}
{\Gamma(\overline{B} \rightarrow \mu^+) + \Gamma(B \rightarrow \mu^- X)}.
\label{eq:absl}
\end{equation}
Assuming CPT symmetry holds, it can be shown\cite{equala} that
$A^b_{\mathrm{sl}} = a^b_{\mathrm{sl}}$.

From the like-sign dimuon sample, after correcting for backgrounds,
the CP-violating asymmetry is found for 9.0~fb$^{-1}$ to be\cite{D0_dimuon2}
$
A^b_{\mathrm{sl}} = -(0.736 \pm 0.172 \thinspace 
{\mathrm{(stat)}}\pm 0.093)\% \thinspace {\mathrm{(syst)}},
$
which is a $3.9\sigma$ deviation from the Standard Model prediction\cite{SMpred}
for CPT-preserving T violation, 
$A^b_{\mathrm{sl}}{\mathrm{(SM)}} = (-2.3^{+0.5}_{-0.6}) \times 10^{-4}$, and 
represents the first evidence for anomalous CP violation in the mixing
of neutral $B$ mesons.  
This asymmetry has contributions from both the CP-violating semileptonic
asymmetry $a^d_{\mathrm{sl}}$  for $B^0_d$ and
$a^s_{\mathrm{sl}}$ for $B^0_s$ oscillations:
$
A^b_{\mathrm{sl}} =
C_d \cdot a^d_{\mathrm{sl}} +
C_s \cdot a^s_{\mathrm{sl}}.
$
The analysis is redone for various conditions on the impact parameters of the 
muons resulting in different values for the coefficients $C_i$ 
allowing for the 
extraction of the specific asymmetries $a^d_{\mathrm{sl}}$  and
$a^s_{\mathrm{sl}}$, albeit with a high degree of correlation:
$a^d_{\mathrm{sl}} = (-0.12 \pm 0.52)\%, 
 a^s_{\mathrm{sl}} = (-1.81 \pm 1.06)\%, \rho_{ds} = -0.7994.$ 
%

A CPT-violating effect in $B$-meson mixing
was predicted some time ago,
\cite{kp,oldcpt}
and the \bsbsb\ system is of particular interest
for studies of CPT violation
because several complete particle-antiparticle oscillations
occur within a meson lifetime
\cite{d0cdf}.
%
A measure of CPT violation is given by the 
inclusive `right-charge' muon charge asymmetry
$\cA^b_{\rm CPT}$
of semileptonic decays of $b$ hadrons,\cite{cpt_preprint}
\begin{equation}
 \cA^b_{\rm CPT} =
 \frac{ \Ga(\bb \to \mu^- X) - \Ga(B \to \mu^+ X) }
 { \Ga(\bb \to \mu^- X) + \Ga(B \to \mu^+ X) } \approx
a^b_{\rm sl}{\mathrm{(SM)}} - A^b_{\rm sl}.
\label{asymmCPT}
\end{equation}
Assuming the only source of T violation
is the SM contribution
$a^b_{\rm sl} {\rm (SM)} = A^b_{\rm sl} {\rm (SM)}$,
we find\footnote{This is the updated number using the result of 
Ref.\ \refcite{D0_dimuon2}.}
$ \cA^b_{\rm CPT} =
0.00785 \pm 0.00300.$
Averaging over sidereal time and the momentum ($\gamma$) spectrum,
sensitivity to the spatial components $(\Delta a^{B_s})_J$ is lost;
however, 
assuming that the only source of CPT violation
comes from \bsbsb\ mixing, the bound 
\begin{equation}
- 3.8 \times 10^{-12} < \dabs_T < 1.1 \times 10^{-11}
\label{result}
\end{equation}
can be set at the 95\% confidence level.
The value of Eq.\ \refeq{result}, documented in Ref.\ \refcite{cpt_preprint},
represents the first sensitivity to CPT violation
in the \bsbsb\ system.

\section{D\O\ exclusive semileptonic $B^0_s$ decays}

In principle, the spatial components $\dabs_J$ could be accessed
by studying the variation of the dimuon like-sign charge asymmetry above as a
function of sidereal time. However, there are complications due to the possibility
of $B^0_d$ CPT asymmetry and the current large uncertainty on 
$\Delta\Gamma_d/\Gamma_d$.\cite{PDG}
Instead, it is better to examine exclusive $B^0_s$ decays, i.e., 
$B^0_s \rightarrow D_s \mu \nu$.  One could fit to a complicated
time-dependent CP or CPT asymmetry. This would require `flavor tagging', i.e., 
determining if the meson was a $B^0_s$ or $\bar{B}^0_s$ at the time of production,
and this costs efficiency.  One can just integrate over decay time since the $B^0_s$
mesons oscillate very quickly and after very small decay times and lengths are `fully mixed', 
i.e., there are equal probabilities of observing a $B^0_s$ or $\bar{B}^0_s$ despite
the flavor at the time of production. 

The relevant semileptonic charge CP asymmetry is therefore:
\begin{equation}
a_{\mathrm{sl}}^s = \frac{N(B_s \rightarrow \bar{B}_s \rightarrow D_s^- \mu^+X) -
     N(\bar{B}_s \rightarrow B_s \rightarrow D_s^+\mu^-X)}{N(B_s \rightarrow \bar{B}_s \rightarrow D_s^- \mu^+X) +
          N(\bar{B}_s \rightarrow B_s \rightarrow D_s^+\mu^-X)}.
\label{eq:assl}
\end{equation}
The D\O\ Collaboration has recently measured\cite{D0_assl} this CP asymmetry in a dataset of integrated luminosity
of 10.4~fb$^{-1}$ resulting in $a_{\mathrm{sl}}^s = [-1.12 \pm 0.74 \thinspace {\mathrm{(stat)}} \pm 0.17 \thinspace
{\mathrm{(syst)}}]\%$. 
This result is consisent with both the
SM value as well as the value extracted from the dimuon like-sign asymmetry. 

The corresponding CPT asymmetry is essentially the {\it same} asymmetry since
the necessary requirement of $B^0_s \rightarrow B^0_s$ and $\bar{B}^0_s \rightarrow \bar{B}^0_s$ 
is identical if the system is fully mixed, only requiring that the observed, raw asymmetry be 
corrected by the fraction of non-oscillated $B^0_s$ decays.  To access all four components of
$\dabs_\mu$, this asymmetry is then  
determined separately in an appropriate number of bins in sidereal time, in this case, optimized as eight bins.  
Information is lost in the binning, so an unbinned analysis is also being performed
in a periodogram\cite{periodogram} that finds the power at a given scan frequency. 
Finally, plans also include determining the  asymmetry in bins of $B^0_s$ meson to probe
for the expected linear variation with $\gamma$ from CPT-violating effects.
Work continues
on this analysis. 

\section*{Acknowledgments}
The D\O\ Collaboration thanks the staffs at Fermilab and collaborating
institutions, and acknowledges support from agencies including the DOE
and NSF (USA).

\end{document}